%% file: main.tex
\documentclass[10pt]{llncs}
\usepackage{graphicx}
\usepackage{epstopdf}
\usepackage[latin1]{inputenc}
\usepackage{amsmath}
\usepackage{amsfonts}
\usepackage{amssymb}
\usepackage{stmaryrd}
\usepackage{color}
\usepackage{bsymb}
\usepackage{floatrow}
\usepackage{pdflscape}

\include{convheader}

\definecolor{dgreen}{rgb}{0,0.5,0}

\newcommand{\KWD}[1]		{{\color{blue}{\texttt{#1}}}}
\newcommand{\KLABEL}[1]		{\mathrm{#1}}
\newcommand{\KIF}			{\KWD{if}}
\newcommand{\KSTOP}			{\KWD{stop}}

\newcommand{\KTHEN}			{\KWD{then}}
\newcommand{\KELSE}			{\KWD{else}}
\newcommand{\KEND}			{\KWD{end}}
\newcommand{\KWHILE}		{\KWD{while}}
\newcommand{\KINV}			{\KWD{invariant}}
\newcommand{\KTH}			{\KWD{theorem}}
\newcommand{\KVAR}			{\KWD{var}}
\newcommand{\KPROCESS}		{\KWD{process}}
\newcommand{\KENV}			{\KWD{environment}}
\newcommand{\KREL}			{\KWD{rely}}
\newcommand{\KGAR}			{\KWD{guar}}
\newcommand{\KIS}			{\KWD{is}}
\newcommand{\KBEGIN}		{\KWD{begin}}
\newcommand{\KELSEIF}		{\KWD{elseif}}
\newcommand{\KASSERT}		{\KWD{assert}~}

\newcommand{\KATOMIC}		{~\KWD{atomic}}
\newcommand{\KSEQ}			{~\KWD{;}~}

\newcommand{\RINT}[1]		{\left \llbracket \begin{array}{@{}l} #1 \end{array} \right \rrbracket}

\newcommand{\qq}[1]			{\mathit{#1}}
\newcommand{\REL}[1]		{\mathsf{#1}}
\newcommand{\SYN}[1]		{\left \langle #1 \right \rangle}
\newcommand{\IDC}[1]		{\left ( #1 \right )^\diamond}
\newcommand{\STATE}[1]		{\operatorname{\Omega}_{#1}}
\newcommand{\defeq}			{\equiv}

\newcommand{\EXAMPLE}[1]	{
$$
\small{
\begin{array}{l}
#1
\end{array}
}
$$
}

\begin{document}
\title{Event-B/SLP}
\author{Alexei Iliasov}

\institute{
      Newcastle University, UK \\
}

\date{}

\maketitle

\begin{abstract}
We show how the event-based notation offered by Event-B may be augmented by algorithmic modelling constructs without disrupting the refinement-based development process.
\end{abstract}

\section{Introduction}

One of the lessons of the DEPLOY project \cite{deploy} is that the industrial application of formal modelling cannot fully succeed by employing just one notation, paradigm and methodology. In the case of Event-B \cite{EventBBook}, one of the language strong sides at the level of abstract design - a simple and versatile notation suitable for a wide range of abstractions - makes the language difficult to apply to concrete designs. Unstructured event-based models often become unwieldy long and verbose when design and implementation decisions are added. 

In this paper we discuss a proposal to extend the event-based notation of Event-B with algorithmic constructs that permit an efficient specification of a large class of concrete designs. 

\begin{figure}[t]
\includegraphics[scale=0.8]{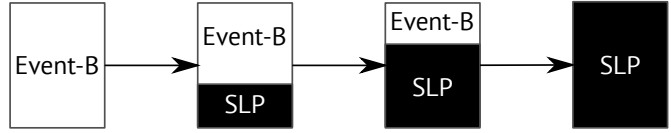} 
\caption{The SLP approach promotes a gradual transition from an event-based to an algorithmic specification.}
\label{fig1}
\end{figure}

Our extension, language SLP (\underline{s}equential composition, \underline{l}oop, \underline{p}arallel composition), is a compact formal modelling notation with strictly defined syntax and semantics. To stay on the same technological platform as Event-B, we define the language semantics as a list of FOL verification conditions. We adopt without changes the mathematical language of Event-B - the part of the notation used to define predicates and expressions. The languages also borrows the notation and the atomicity assumption of Event-B substitutions. 

Rather than a replacement or a simple superposition of algorithmic and event styles we propose to have a seamless connection between Event-B and SLP where a high-level event specification is gradually transformed into an algorithmic specification with explicit concurrency and control flow (see Fig. \ref{fig1}). 

The defining difference between SLP and Event-B is that the latter is data-driven while the former features explicit control flow for sequential computation and units of concurrency for concurrent computations. This requires a departure from a flat machine structure, apt for inductive reasoning but often onerous in practice for large models, to a hierarchical model with nested naming scopes delineating verification concerns.

\section{Syntax}

An SLP model is made of the following three main parts. The first, taken verbatim from Event-B, provides definitions of types, axiom, variables, invariants and theorems. This part may also contain Event-B events in the case of a mixed Event-B/SLP model.

The second part is the definition of \emph{environment} activities. In SLP, we take a view that actions performed by an environment must be explicitly defined as such. This is not just a syntactic notion - SLP offers differing refinement rules (not discussed in this paper) for environment and system activities. 

The final part is the definition of the behaviour of a modelled system. It takes the form of a list of so-called \emph{process} definitions - concurrent units of system behaviour. The body of a process is defined by a succession of \emph{atomic state updates} (substitutions, in the Event-B terminology) connected by the typical algorithmic control structures - sequential composition, \emph{if} and \emph{loop}. A process body runs in an infinite loop until it explicitly executes a termination command. 

The processes of a system and environment activities execute concurrently. They interact by reading and writing shared (global) variables. A system process may also define its private (local) variables to deal with computations that do not need to be exposed to environment or other system processes. For a given process, the \emph{universe} of the process is the set of all other processes and all the environments.

The following is the top-level structure of an SLP specification:
$$
\begin{array}{rcl}
slp 		   & := & \SYN{\qq{invdef}}^* \\
			   &    & \SYN{environment}^* \\ 
			   &    & \SYN{process}^+ ~ ; \\
\SYN{\qq{invdef}}	   & := & (\KINV ~|~ \KTH) ~\SYN{\qq{label}}: ~\SYN{predicate} ~; \\
\end{array}
$$

\noindent
To simplify the presentation, we omit the declaration of constants and sets while variable declarations are deduced from invariants. \footnote{Note that this is our preferred \emph{concrete} syntax. The abstract syntax for these elements is exactly that of Event-B} All the variables defined at the global level are seen by system and environment processes. These should be the variables used to model input/output between the environment and system components.
%
%
%
Like in Event-B, we split invariant conditions to label and partition invariant preservation conditions. 

An environment is a labelled pair of a rely and guarantee predicates. Like invariants, rely and guarantees are labelled.

$$
\begin{array}{rcl}
\SYN{environment} & := & \KENV ~ \SYN{\qq{label}} ~ 
		\SYN{\qq{reldef}}^* ~
		\SYN{\qq{gardef}}^* ~ \KEND; \\
\SYN{\qq{reldef}}	   & := & \KREL ~\SYN{\qq{label}}: ~\SYN{predicate} ~; \\
\SYN{\qq{gardef}}	   & := & \KGAR ~\SYN{\qq{label}}: ~\SYN{predicate} ~; \\
\end{array}
$$

The following is an example of an environment describing the behaviour of a temperature sensor. The environment may update value of $t$ (current temperature) by changing it in some small increments defined by constant $\Delta$. A rely predicate is omitted and assumed to be $\btrue$.

\EXAMPLE{
\KENV ~ \KLABEL{temp\_sensor}\\
\qquad \KGAR ~\KLABEL{guar1} ~\KIS ~ t' \in t - \Delta \upto t + \Delta \\	
\KEND
}

A system activity, called a process, follows the template of an environment but may also define local variables and concrete behaviour specification. 
$$
\begin{array}{rcl}
\SYN{process} & := & \KPROCESS ~ \SYN{\qq{label}} ~ 
			\SYN{\qq{reldef}}^* ~
			\SYN{\qq{gardef}}^* ~
			\SYN{\qq{invdef}}^* ~
			\SYN{block}? ~
			\KEND; \\
\end{array}
$$

\noindent
Informally, the body of a process is the implementation that is shown to tolerate the interference defined by the process rely and satisfy the obligation of the process guarantee. In an extreme case of a solipsistic process there may be no rely and guarantee predicates so that the process has no specific obligations to its universe. Such a process specifies a sequential algorithm that runs till completion without any interaction.

Continuing the theme of the sensor example, with the syntax discussed, we can already define a small but meaningful specification. The temperature sensor $t$ belongs to the environment while the system controls the heater modelled by variable $heater$:

\EXAMPLE{
\KINV ~\LABEL{temp} ~ t \in \intg \\
\KINV ~\LABEL{heater} ~ h \in \Bool \\
\KENV ~ \LABEL{temp\_sensor}\\
\qquad \KGAR ~\KLABEL{guar1} ~\KIS ~ t' \in t - \Delta \upto t + \Delta \\	
\KEND \\
\KPROCESS ~ \KLABEL{heater\_control}\\
\qquad \KREL ~\LABEL{rel1}  ~ t \in \mathrm{SAFE\_TEMP} \\
\qquad \KGAR ~\LABEL{guar1} ~ t > \mathrm{TEMP\_HIGH} \land h = \True \limp h' = \False \\
\qquad \KGAR ~\LABEL{guar2} ~ t < \mathrm{TEMP\_LOW} \land h = \False \limp h' = \True \\
\KEND
}

\noindent
There may be any number of $\KENV$ and $\KPROCESS$ parts. One may, for instance, add an alarm process to detect an abnormal temperature range.

\EXAMPLE{
\KINV ~\LABEL{alarm} ~ alarm \in \Bool \\
\KPROCESS ~ \KLABEL{alarm\_control}\\
\qquad \KGAR ~\LABEL{guar1} ~ alarm' = \bool(t \notin \mathrm{SAFE\_TEMP}) \\
\KEND
}

\noindent
Note that the rely of $\mathrm{heater\_control}$ is not always satisfied by the sensor behaviour. A system process is temporarily disabled if its rely is broken by an environment. A process, however, may not violate the rely of another process or an environment.

The body of a process describes how the activity defined by its guarantee predicate is realised. The following operators are used to build the body of a process:

$$
\begin{array}{lll}
\SYN{block} & := & \SYN{action} \KSEQ \SYN{block} ~; \\
\SYN{action} & := & \SYN{statement} \KATOMIC? ~ \SYN{refines}? ~ \SYN{with}? \\
\SYN{statement} & := & \SYN{substitution} ~|~ \SYN{\qq{if}} ~|~  \SYN{loop} ~|~ \SYN{\qq{begin\_end}} ~|~ \SYN{\qq{assert}} ~|~ \KSTOP ~; \\
\SYN{\qq{if}} & := & \KIF~ \SYN{predicate} ~ \KTHEN~ \SYN{block} \\
			  &    & (\KELSEIF~ \SYN{predicate} ~ \KTHEN~ \SYN{block})^* \\
			  &    & (\KELSE~ \SYN{block})?~ \KEND ~; \\
\SYN{loop}		   & := & \KWHILE ~ \SYN{predicate} \\
			   &	& \SYN{\qq{invdef}}^* \\
			   &	& \KVAR ~ \SYN{expression} \\
			   &	& \KTHEN~ \SYN{block} ~ \KEND ~; \\
\SYN{\qq{begin\_end}}	   & := & \KBEGIN ~ \SYN{\qq{invdef}}^* ~ \SYN{block} ~ \KEND ~; \\
\SYN{\qq{assert}}    & := & (\KASSERT~ (\SYN{\qq{label}}:)? ~ \SYN{predicate})^+ 
\end{array}
$$

\noindent
Most of the syntax is self explanatory. The $\KSTOP$ statement terminates a process; $\KASSERT p$ asserts the truth of $p$; $\SYN{substitution}$ and $\SYN{expression}$ are Event-B substitution and expression elements (see Rodin Deliverable D7 \cite{D7} for concrete definitions). Block $\KBEGIN\_\KEND$ defines the scope of visibility for local variables. Elements $\KATOMIC$, $\SYN{refines}$ and $\SYN{with}$ are used to define the refinement relationship between SLP models but are not discussed in this paper.

A trivial implementation of $\mathrm{heater\_control}$ retells the implications in the process guarantee as an \emph{if} statement:

\EXAMPLE{
\KPROCESS ~ \KLABEL{heater\_control}\\
\qquad \KREL ~\LABEL{rel1} ~ t \in \mathrm{SAFE\_TEMP} \\
\qquad \KGAR ~\LABEL{guar1} ~ t > \mathrm{TEMP\_HIGH} + \delta \land h = \True \limp h' = \False \\
\qquad \KGAR ~\LABEL{guar2} ~ t < \mathrm{TEMP\_LOW} - \delta  \land h = \False \limp h' = \True \\
\qquad \KIF ~ t > \mathrm{TEMP\_HIGH} + \delta \land h = \True ~\KTHEN \\
\qquad \qquad \LABEL{act1} ~ h' \bcmeq \False \\
\qquad \KELSEIF ~ t < \mathrm{TEMP\_LOW} - \delta  \land h = \False ~ \KTHEN\\
\qquad \qquad \LABEL{act2} ~ h' \bcmeq \True \\
\qquad \KEND \\
\KEND
}

\input{semantics}

\input{b2slp}

\section{Small Example}

\input{example}

\section{Conclusion}

The implementation language of B-Method, B0 \cite{BBook} is one of the inspirations for this works. There are, however, important differences in both aims and techniques employed: B0 allows a modeller to write more detailed bodies of abstract operations using the concepts from programming languages. In contrast, in SLP, the main development technique is an aggregation of several abstract events into a body of a process. This means that a data-driven design of Event-B may be refined into an algorithmic design whereas in B0 it would have to remain data-driven at the top level. Equally important is an explicit treatment of concurrency that becomes more and more relevant topic in embedded systems design. We use rely/guarantee \cite{Jones83a} approach to model cooperation of concurrent processes via shared variables. 

Event-B is rather obviously lacking in means of control flow specification. One solution is the integration of two narrowly specialised two notation, i.e., CSP$\|$B that combines B and CSP \cite{Treharne}. Another is extension of the basic notation with means to explicitly define control flow, i.e., the Flow plug-in for Rodin \cite{Serene11}. In this paper we followed a different direction with a premise that a deficiency of a notation in a certain area is best rectified by coming up with a new notation.

This leads us to the following crucial point: to make Event-B applicable in any given problem domain it may be necessary to (1) design a specialised concrete syntax exposing Event-B method in a way tailored to the problem domain (for example, a graphical notation like the one offered by UML-B \cite{UMLB}) and (2) devise a specialised notation and refinement rules for concrete designs, like the one shown in this paper. The use of Event-B for an abstract design puts a development on a solid and well-studied platform. But concrete designs incorporating implementation decision must  offer the concepts, terminology and structuring principles already employed and recognised in the target problem domain. In this sense, the language defined in this paper is merely a technological demonstration that such a direction is viable. 

\bibliographystyle{plain}
\bibliography{allrefs2}

\end{document}

%% file: convheader.tex
\newcommand{\kG}[1]					{{\color{blue}{\texttt{#1}}}}
\newcommand{\kB}[1]					{\mathsf{#1}}

\newcommand{\kwSYSTEM}			{machine}

\newcommand{\kwREFINEMENT}		{refinement}
\newcommand{\kwVARIABLES}		{variables}
\newcommand{\kwINVARIANT}		{invariant}
\newcommand{\kwINITIALISATION}	{initialisation}
\newcommand{\kwEVENTS}			{events}

\newcommand{\kwMEND}				{end}

\newcommand{\kwTHEN}			{then}
\newcommand{\kwWHEN}			{when}

\newcommand{\kwEND}				{end}
\newcommand{\kwBEGIN}			{begin}

\newcommand{\kwREFINES}			{refines}
\newcommand{\LABEL}[1]			{\mathrm{#1}:~}

\newcommand{\EventBSystem}[2] {
	{\footnotesize{
	$
	\begin{array}{@{}l}
		\kG{\kwSYSTEM} ~ \kB{#1} \\
		\quad \begin{array}{@{}l} #2 \end{array} \\
		\kG{\kwMEND}
	\end{array}
	$
	}}
}

\newcommand{\EventBRefinement}[2] {
	{\footnotesize{
	$
	\begin{array}{@{}l}
		\kG{\kwREFINEMENT} ~ \kB{#1} \\
		\quad \begin{array}{@{}l} #2 \end{array} \\
		\kG{\kwMEND}
	\end{array}
	$
	}}
}

\newcommand{\EventBRefines}[1] {
	\kG{\kwREFINES} ~ \kB{#1} \\
}

\newcommand{\EventBVarsC}[1] {
	\kG{\kwVARIABLES} ~ #1 \\
}

\newcommand{\EventBInv}[1] {
	\kG{\kwINVARIANT} \\
	\quad \begin{array}{l} #1 \end{array} \\
}

\newcommand{\EventBInvC}[1] {
	\kG{\kwINVARIANT} ~ #1 \\
}

\newcommand{\EventBInitC}[1] {
	\kG{\kwINITIALISATION} ~ #1 \\
}

\newcommand{\EventBEvents}[1] {
	\kG{\kwEVENTS} \\
	\quad \begin{array}{lll} #1 \end{array} \\
}

\newcommand{\EventBWhenC}[3] {
   \kB{#1} & = 	& \kG{\kwWHEN} ~#2~ \kG{\kwTHEN} ~#3~ \kG{\kwEND} \\
}

\newcommand{\EventBBeginC}[2] {
   \kB{#1} & = 	& 	\kG{\kwBEGIN} ~ #2 ~ \kG{\kwEND} \\
}

%% file: semantics.tex
\subsection{Semantics}

Similar to Event-B, the semantics of SLP is given as a list of verification conditions called proof obligations. We discuss only the consistency conditions showing that the SLP part of an Even-B/SLP model does not violate invariants and introduce deadlocks and divergences. Informally, the purpose of consistency proof obligations is to establish the following three facts:

\begin{itemize}
\item when control is passed to a statement, the state update defined by the statement may take place;
\item any statement does not take the system outside of the safety invariant bounds;
\item a statement eventually terminates.
\end{itemize}

We begin by cataloguing the major syntactic elements of a specification. The following are coming from Event-B and are shared between Event-B and SLP models: constants $c$, carrier sets $s$, axioms $P(c, s)$, global variables $v$ and invariant $I(c, s, v)$.

There are elements specific to SLP. Taking the viewpoint of a substitution $S$ located somewhere in the body of a process, they are: the rely $R(c, s, v, v')$ and guarantee $G(c, s, v, v')$ of a current process; process variables $u$ (must be distinct from $v$); process invariant $T(c, s, v, u)$; variables defined in enclosing $\KBEGIN \dots$ and $\KWHILE \dots$ blocks, $\mathbf{w} = \{w_1, \dots, w_i\}$ (all distinct); $\KBEGIN \dots$ and $\KWHILE \dots$ block invariants $B_i(c, s, v, u, w_1, \dots, w_i)$; assertion predicate $A(c, s, v, u, \mathbf{w})$ expressed directly in a preceding $\KASSERT$ or derived from other kind of a preceding statement; and, finally, the substitution itself - $S(c, s, v, u, \mathbf{w}, v', u', \mathbf{w}')$. 

The following shorthand is used to identify syntactic element in the context of substitution $S$. Assume that $S$ is contained inside $i$ nested blocks $\KBEGIN/\KWHILE$ that define some local variables $u$ and $\mathbf{w}$. In the scope of $S(\dots)$ the actual invariant is $\mathcal{I}_i$, as defined below. The invariant defines the state space $\STATE{i}$ on which the update defined by $S$ takes the effect: $\{z \mid S(z)\} \subseteq \STATE{i} \times \STATE{i}$.

$$
\begin{array}{lll}
\mathcal{I}_i = \left (
\begin{array}{l}
P(c, s) \\
I(c, s, v) \\ 
T(c, s, v, u) \\ 
\bigwedge_{j \leq i} B_j(c, s, v, u, w_1, \dots, w_j) \\ 
\end{array} 
\right ) 
& \qquad &
\begin{array}{l}
\mathcal{A} = A(c, s, v, u, \mathbf{w}) \\
\mathcal{S} = S(c, s, v, u, \mathbf{w}, v', u', \mathbf{w}') \\
\STATE{i} = \{z \mid \mathcal{I}_i(z)\} \\
 \STATE{i}^\surd = \STATE{i} \cup \{\surd\}
\end{array}
\end{array}
$$

\noindent
Extended state $\STATE{i} \cup \{\surd\}$ adds a termination symbol $\surd$ from which no continuation is possible. Globally, the set of all names spaces forms a tree such that the state of an inner wholly contains the state of outer space: $\STATE{0} \subseteq \STATE{1} \subseteq \dots \subseteq \STATE{n}$ where $\STATE{0}$ is the state of a name space of containing just global variables and $\STATE{n}$ is the state of some current block within the body of a process.

To define verification conditions, we convert SLP statements into relations describing the connection between previous and next states. All the partial state update relations are treated as \emph{guarded relations} (i.e., never applied outside of their domain) and loops are required to terminate to ensure \emph{total correctness}. We write $\REL{II}$ meaning $[\mathcal{I}_i]$, $\REL{I}$ meaning $[I]$, $\REL{A}$ for $[\mathcal{A}]$ and so on.

$$
\begin{array}{rcl}
\RINT{\dots}_i & \in & \STATE{i} \rel \STATE{i}^\surd \\ 
\RINT{\KSTOP}_i & := & \STATE{i} \times \{\surd\} \\
\RINT{\KASSERT~ p }_i & := & \id([p] \cap \STATE{i}) \\
\RINT{a \KSEQ \KASSERT p \KSEQ b}_i & := & \IDC{[p] \domres \RINT{b}_i} \\
\RINT{a \KSEQ b}_i & := & \RINT{b}_i \circ \RINT{a}_i \\
\RINT{s_1 \| \dots \| s_n}_i & := & \RINT{s_1}_i \cup \dots \cup \RINT{s_n}_i \\

\RINT{u \bcmeq E(v)}_i & := & \{u \mapsto E(v)\}^{\diamond} \\
\RINT{u \bcmin E(v)}_i & := & \{u \mapsto u' \mid u' \in E(v)\}^{\diamond} \\
\RINT{u \bcmsuch E(v, v')}_i & := & \{u \mapsto u' \mid E(v, v')\}^{\diamond} \\
\RINT{
\KIF~ c_0 ~ \KTHEN~ b_0 \\
\KELSEIF~ c_1 \KTHEN ~ b_1 \\
\dots \\
\KELSEIF~ c_k \KTHEN ~ b_k \\
\KELSE~ b_e~ \KEND
}_i\!
& := &
\begin{array}{l}
\IDC{
(\mathbf{c}_0 \domres b_0) \cup (\mathbf{c}_1 \domres b_1) \cup \dots \cup (\mathbf{c}_k \domres b_e)
}
 \\
\text{where} ~ \mathbf{c}_i = c_i \setminus (\bigcup_{j \in 0 \upto i-1} c_j)
\end{array}\\

\RINT{
\KWHILE ~ c \\
\KINV ~ LI \\
\KVAR ~ V \\
\KTHEN~ b ~ \KEND ~
}_i\!
& := &
\RINT{\KASSERT \neg c \land LI \land \mathsf{trm}\left (C, LI, V, \RINT{b}_{i+1} \right )}_i \\

\RINT{
\KBEGIN \\
\KINV ~ BI\\
b  \\
\KEND
}_i\!
& := &
\IDC{ [BI] \domres \RINT{b}_{i+1} \cap \left ( \STATE{i} \times \STATE{i}  \right ) } \\
\end{array}
$$

\noindent
Operator $r^{\diamond}$ extends a relation $r \subseteq \STATE{i} \times \STATE{i}^\surd$ to a total relation $r' \subseteq \STATE{i} \trel \STATE{i}^\surd$ so that mappings not covered by $r$ are taken from $\id(\STATE{i})$:
$r^{\diamond} = \{ x\mapsto y \mid x \mapsto y \in r \lor x \mapsto y \in \id(\STATE{i}) \setminus r\} = \id(\STATE{i}) \ovl r$. Also, as a shorthand, for some predicate $x \in \STATE{I} \tfun \Bool$ we write $[x]$ to mean a set of elements satisfying $x$: $[x] = \{e \mid x(e)\}$.

In the definition of a loop, $V \in \STATE{i} \tfun \nat$ is a loop variant and $\mathsf{trm}$ is a termination condition expressing that the variant value is decreased by each loop iteration: $\mathsf{trm}(C, LI, V, b) \defeq \forall x, y \cdot x \in V \left [b[st] \right ] \land y \in V\left [ st \right] \limp x < y$, where $st \equiv \REL{I} \cap [LI \land c]$.

Note the two rules for sequential composition. The $a \KSEQ b$ case defines the conventional sequence-to-relational-join rule. The preceding rule (of a higher precedence) makes the sequential composition 'forgetful' when an assertion is placed between two statements: the information about previous statements is dropped and the focus is placed on the last statement and the preceding assertion. One reason for this is that a chain of substitutions may lead to a large and intractable set of hypothesis preventing efficient automated proof and introduce an undesirable interdependency between substitutions where a change in one substitution could invalidate proofs done for successive substitutions. An assertion breaks such a chain making the proof context smaller. Another reason, specific to our technique of refining Event-B into SLP, is the use of assertions to prove that the set of enabling states of a refined substitution does not grow larger in a refined model. 

The following is a list of the more important proof obligation, given, for brevity, in a relational form. 

\paragraph{Well-definedness} SLP mirrors the Event-B approach of proving that each partial relation is well-guarded. In other words, we prove that a relation defined by statement $a$ may be applied to a current state: $(\REL{II} \cap \REL{A}) \domres \RINT{a} \neq \emptyset$.

\paragraph{Feasibility of rely} The rely must not contradict an invariant: $\REL{I} \domres \REL{R} \subseteq \REL{I} \times \REL{I}$.

\paragraph{Closure of rely} Conditions involving rely invariably require tolerating any number of rely iterations. To simplify corresponding proof obligations we insists that a rely relation $\REL{R}$ is reflexively and transitively closed: $\id(\STATE{i}) \subseteq \REL{R} \land \REL{R} \circ \REL{R} \subseteq \REL{R}$.

\paragraph{Invariant preservation} Invariant properties of model variables are assumed to hold before every substitution. It must be proven that all invariants known in the scope of a substitution are re-established by the substitution: $\RINT{a}[\REL{II} \cap \REL{A}] \subseteq \REL{II}$.

\noindent
Not that when statement $a$ is located in the body of a loop $\REL{II}$ also includes the loop invariant. 

\paragraph{Variant} A loop variant is based on the same principles as Event-B variant and is embedded into the rule converting a loop into a relational form.

\paragraph{Establishing guarantee} A substitution executed by a process must agree with a process guarantee. Formally, any state update would be covered by a 'promise' expressed in the guarantee: $ (\REL{II} \cap \REL{A}) \domres \RINT{a} \subseteq \REL{G}$.

\paragraph{Establishing assertion} An asserted condition $A_n$ must be implied by a previous assertion or a statement. We must take into the account the fact that between previous and current statements the universe might have changed its state. For this, the latest locally known state is 'blurred' by the rely condition of a process. 

\begin{itemize}
\item if two assertions follow each other then the second must be contained in the first:
	$
	\left ( \REL{A} \domres \REL{R} \right )[\REL{II}] \subseteq \REL{A}_n
$;	
	
\item otherwise, if an assertion is preceded by a substitution, the preceding substitution after-state must imply the assertion:
$
\left (  \REL{R} \circ \RINT{a}  \right )[\REL{II}] \subseteq \REL{A}_n
$;

\item otherwise, an assertion must be established by an invariant:
	$
	\REL{R}[\REL{II}] \subseteq \REL{A}_n
$.	
	
\end{itemize}

\paragraph{Process compatibility} All non-environment processes must be compatible w.r.t. their rely/guarantee conditions: $\REL{I} \domres \REL{G}_A \subseteq \REL{R}_B$.

%% file: b2slp.tex
\section{From Event-B to SLP}
SLP is not a standalone formalism and is meant to complement the Event-B notation when one needs to obtain a detailed design expressed in terms of parallel processes and algorithmic constructs. Thus, there is always a stage when a pure Event-B specification undergoes a transformation into an Event-B/SLP specification. 

One simple case of Event-B to Event-B/SLP refinement is introducing environments and processes operating on new variables. In a general case, the Event-B part is refined to make use of new variables so that there is an information flow between the two parts. Naturally, there are no specific proof obligations for this case: one only needs to discharge the consistency conditions.

A more interesting situation is the \emph{replacement} of Event-B events with SLP constructs. Of all possibilities, we shall only consider the simplest one: refinement of a set of events by \emph{new} (rather than existing) environments and processes. 

\paragraph{New environment (process)} A new SLP environment (process) may be defined to refine one or more abstract Event-B events; refined events disappear from a model. The relevant proof obligation is that a process guarantee is contained in the behaviour of refined events: $(\REL{I} \cap \REL{R}) \domres \REL{G} \subseteq [e_1]_R \cap \dots \cap [e_n]_R$.

\paragraph{New concrete process} A sub-set of machine events may be refined into a process with a body. We focus on a simpler case when this is done in a single refinement step. Without loss of generality, we consider the case of refinement where substitutions of a process body coincide exactly with substitutions of refined events, in other words, a refinement that forms a process from events without any further behavioural or data refinement that may take place in following refinement steps. 

Let $E$ be the set of events of machine $M$ describing the behaviour of a prospective process $P$ and $tr(M) \upharpoonright E$ be the machine traces limited to events $E$. Let $tr(P)$ be a set of traces of a new process in terms where each trace element is the list of labels of parallel substitution parts. It is easy to define a mapping $f$ from the alphabet of $tr(P)$ to set $E$ (it is not necessarily a one-to-one mapping but this does not pose problems). If one can prove that $f(tr(P)) \subseteq tr(M) \upharpoonright E$ and, separately, that process $P$ does not introduce new divergences then process $P$ is declared to refine events $E$. We have previously shown how to convert a statement of the form $f(tr(P)) \subseteq tr(M) \upharpoonright E$ into a list of FOL theorems \cite{Serene11,FlowAdaEurope}.

%% file: example.tex
We illustrate the Event-B/SLP hybrid modelling by showing a simple case of Event-B to SLP refinement. The model computes the greatest common devisor (GCD) of two numbers. Function $\mathrm{gcd} \in \nat \times \nat \tfun \nat$ axiomatically satisfies the following properties:

$$
\begin{array}{l}
\mathrm{axm1}: \forall a, b \cdot a, b \in \nat \land a > b \limp 
		\mathrm{gcd}(a, b) = \mathrm{gcd}(a - b, b) \\
\mathrm{axm2}: \forall a, b \cdot a, b \in \nat \land b > a \limp 
		\mathrm{gcd}(a, b) = \mathrm{gcd}(a, b - a) \\
\mathrm{axm3}: \forall a \cdot a \in \nat \limp \mathrm{gcd}(a, a) = a
\end{array}
$$

\noindent
At an abstract level, one may use the constant function $\mathrm{gcd}$ to compute the result in one step:

\EventBSystem{gcd0}{
	\EventBVarsC{r, x1, x2}
	\EventBInvC{r \in \nat \land x1 \in \nat \land x2 \in \nat}
	\EventBInitC{r \bcmin \nat ~\|~ x1 \bcmin \nat ~\|~ x2 \bcmin \nat}
	\EventBEvents{
		\EventBBeginC{gcd}{r \bcmeq \mathrm{gcd}(x1 \mapsto x2)}
	}
}

\noindent
Variables $x1$ and $x2$ serve as input values and $r$ holds the result. The following is a typical Event-B refinement based on the unfolding of an atomic abstract step into a sequence of concrete computations. 

\EventBRefinement{gcd1a}{
	\EventBRefines{gcd0}
	\EventBVarsC{r, x1, x2, y1, y2, pc}
	\EventBInv{
		y1 \in \nat \land y2 \in \nat \\	
		pc \in 1 \upto 5 \\
		pc = 2 \limp \mathrm{gcd}(x1 \mapsto x2) = \mathrm{gcd}(y1 \mapsto x2) \land y1>0 \land x2>0 \\
		pc = 3 \limp \mathrm{gcd}(x1 \mapsto x2) = \mathrm{gcd}(y1 \mapsto y2) \land y1>0 \land y2>0 \\
		pc = 4 \limp \mathrm{gcd}(x1 \mapsto x2) = gcd(y1 \mapsto y2) \land y1>0 \land y2>0 \\ 
	}
	\EventBInitC{\dots ~\|~ y1 \bcmin \nat ~\|~ y2 \bcmin \nat ~\|~ pc \bcmeq 1}
	\EventBEvents{
		\EventBWhenC{copy1}{pc = 1}{y1 \bcmeq x1 ~\|~ pc \bcmeq 2}
		\EventBWhenC{copy2}{pc = 2}{y2 \bcmeq x2 ~\|~ pc \bcmeq 3}
		\EventBWhenC{sub1}{y1 > y2 \land pc \in \{3, 4\}}{y1 \bcmeq y1 - y2 ~\|~ pc \bcmeq 4}
		\EventBWhenC{sub2}{y2 > y1 \land pc \in \{3, 4\}}{y2 \bcmeq y2 - y1 ~\|~ pc \bcmeq 4}			
		\EventBWhenC{gcd}{y1 = y2 \land pc = 4}{r \bcmeq y1}
	}
}

Events $\kB{sub1}$ and $\kB{sub2}$ form the body of a loop. An auxiliary variable $pc$ is used to simulate control flow; variables $x1, x2, y1, y2$ are introduced to describe the concrete computation steps. Note how the after state of each event is encoded in model invariant. The repeating template $v = C \limp \dots$ in invariants is an  indicator that an event-based specification is used to simulate concrete control flow.

The SLP version of the same refinement step is given below. Here we have an explicit loop construct containing a two-branch \emph{if} that makes for a more concise specification without the need to propagate state properties via an invariant.

\EventBRefinement{gcd1b}{
	\EventBRefines{gcd0}
	\EventBVarsC{r, x1, x2, y1, y2}
	\EventBInvC{y1 \in \nat \land y2 \in \nat}
	\EventBInitC{\dots ~\|~ y1 \bcmin \nat ~\|~ y2 \bcmin \nat}
	\KPROCESS~\mathit{main} \\
	\quad y1 \bcmeq x1 ~\|~ y2 \bcmeq x2 \KSEQ \\
	\quad \KWHILE ~ y1 \neq y2 ~\KTHEN \\
	\quad \KINV ~ gcd(x1 \mapsto x2)=gcd(y1 \mapsto y2) \land y1>0 \land y2>0 \\
	\quad \quad \KIF ~ y1 > y2 ~ \KTHEN ~ y1 \bcmeq y1 - y2 \\
	\quad \quad \KELSEIF ~ y2 > y1 ~ \KTHEN ~ y2 \bcmeq y2 - y1 ~ \KEND\\
	\quad \KEND \KSEQ \\
	\quad r \bcmeq y1 \\
	\KEND
}

\noindent
Essential to the proof of refinement is the last case of sequential composition where control is passed from a loop to an assignment saving the final result. The relational interpretation of the loop asserts the loop invariant and the negation of the loop condition which immediately give that $r=y1=gcd(x1 \mapsto x2)$.